# VoxCap: FFT-Accelerated and Tucker-Enhanced Capacitance Extraction Simulator for Voxelized Structures

Mingyu Wang, *Student Member, IEEE*, Cheng Qian, Jacob K. White, *Fellow, IEEE*, and Abdulkadir C. Yucel, *Senior Member, IEEE*

*Abstract*—VoxCap, a fast Fourier transform (FFT)-accelerated and Tucker-enhanced integral equation simulator for capacitance extraction of voxelized structures, is proposed. The VoxCap solves the surface integral equations (SIEs) for conductor and dielectric surfaces with three key attributes that make the VoxCap highly CPU and memory efficient for the capacitance extraction of the voxelized structures: (i) VoxCap exploits the FFTs for accelerating the matrix-vector multiplications during the iterative solution of linear system of equations arising due to the discretization of SIEs. (ii) During the iterative solution, VoxCap uses a highly effective and memory-efficient preconditioner that reduces the number of iterations significantly. (iii) VoxCap employs Tucker decompositions to compress the block Toeplitz and circulant tensors, requiring the largest memory in the simulator. By doing so, it reduces the memory requirement of these tensors from hundreds of gigabytes to a few megabytes and the CPU time required to obtain Toeplitz tensors from tens of minutes (even hours) to a few seconds for very large scale problems. VoxCap is capable of accurately computing capacitance of arbitrarily shaped and large-scale voxelized structures on a desktop computer. VoxCap's accuracy, efficiency, and capability are demonstrated through capacitance extraction of various large-scale structures, including the parallel meander lines discretized by more than a hundred million panels and analyzed on a commodity desktop computer.

## I. Introduction

Recent developments in three-dimensional (3D) printing technology have allowed the designers to produce their own prototypes conveniently and reduce the prototype development time dramatically. Today's 3D printers build the integrated circuits/packages/components voxel by voxel (i.e., cube by cube) [1], [2]. Furthermore, virtual fabrication environments for the process modeling of the semiconductors and microelectromechanical systems (MEMS) leverage voxels [3], [4]. By using these environments, the designers can iteratively explore their designs; they can alter their designs and check whether their designs meet certain design specifications. During the design process of 3D printed integrated circuits as well as the semiconductor devices and MEMS via the virtual fabrication environments, the designers are in need of fast and accurate parameter extraction simulators that can be operated on the structures discretized by voxels (i.e., voxelized structures). Unfortunately, the current literature does not have enough studies on the development of such simulators: An inductance extraction simulator for voxelized structures, called VoxHenry [5], was recently proposed. That said, there exists no study on the capacitance extraction for the voxelized structures so far.

The literature abounds with the capacitance extraction simulators based on finite-difference [6]-[8], finite-element [9]-[11], integral equation [12]-[20] methods. While these methods are no panacea, none of them is developed for / directly applicable to the voxelized structures and can readily extract capacitances by only using the voxel coordinates provided from the virtual fabrication environment. When used for voxelized structures, all current simulators require the users' intervention such as the inclusion of artificial absorbing boundaries or generation of complicated mesh. Furthermore, existing integral equation simulators (leveraging multipole expansions [19], pre-corrected fast Fourier transform (FFT) [20], or H/$H^2$ matrices [15]) were developed to be operated on the meshed structures with the non-uniform elements, which are not (necessarily) residing on a structured grid. These simulators do not exploit the distinct features of the voxelized structure with uniform elements residing on a structured grid. Thereby, they are not sufficiently efficient when applied to the voxelized structures. Aside from that, these simulators cannot be easily coupled with the VoxHenry to extract impedances of the voxelized structures. To this end, a capacitance extraction simulator that can exploit all distinct features of voxelized structures, be used in conjunction with voxel-based virtual fabrication environments without users' intervention, and be easily coupled with the VoxHenry simulator is called for.

In this study, a capacitance extraction simulator for voxelized structures, called VoxCap, is proposed. The VoxCap

Manuscript received June 20, 2020. This work was supported by Ministry of Education, Singapore, under grant AcRF TIER 1-2018-T1-002-077 (RG 176/18), and the Nanyang Technological University under a Start-Up Grant. (*Corresponding author: Abdulkadir C. Yucel.*)

M. Wang, C. Qian and A. C. Yucel are with the School of Electrical and Electronic Engineering, Nanyang Technological University, Singapore 639798. (e-mails: mingyu003@e.ntu.edu.sg, cqian@ntu.edu.sg, acyucel@ntu.edu.sg).

J. K. White is with the Department of Electrical Engineering and Computer Science, Massachusetts Institute of Technology, Cambridge, MA 02139, USA. (email: white@mit.edu)

solves the surface integral equations (SIEs) for the conductor and dielectric surfaces after discretizing the charges on these surfaces via piecewise constant basis functions and obtaining a linear system of equations (LSE) via Galerkin testing. It leverages the following three features, which make the VoxCap highly CPU and memory-efficient while solving the LSE and obtaining the capacitances of voxelized structures:

1. The VoxCap exploits the FFTs to accelerate the matrix-vector multiplications (MVMs) during the iterative solution of LSE.
2. The VoxCap uses an effective and memory-efficient block-diagonal-diagonal preconditioner to ensure the fast convergence of the iterative solution.
3. The VoxCap leverages Tucker decompositions to dramatically reduce the memory requirement of the block circulant tensors as well as the CPU time required to obtain Toeplitz tensors.

All these features make the contribution of this study threefold, as described below. The FFTs have been widely applied to integral equation simulators to accelerate the MVMs by exploiting the translationally-invariance property of the integral kernels sampled on the structured grids [5], [21]-[24]. Although it was applied to the 2D static problems in [25], there exists no study providing the implementation of the FFT acceleration for the capacitance extraction of 3D voxelized structures in the literature, to the best of our knowledge. Along with the FFT accelerations, a memory-efficient preconditioning technique hybridizing block-diagonal and diagonal preconditioners is proposed in this study by showing its effectiveness compared to those of traditional diagonal and block-diagonal preconditioners. Moreover, Tucker decompositions are proposed for the first time in this study to compress the block circulant and Toeplitz tensors of static integral kernels, which require the largest memory in the simulator. The proposed Tucker decomposition reduces the memory requirement of these tensors more than four orders of magnitude (from tens/hundreds of gigabytes to megabytes). The Tucker-compressed Toeplitz tensors are obtained *once and for all* and stored on hard-disk during the installation stage of the simulator. During the setup stage of each execution of the VoxCap, the compressed Toeplitz tensors are read from the hard disk in seconds, resized/updated by accounting for the sizes of computational domain and voxels, and used to obtain the block circulant tensors. Doing so reduces the setup stage of the proposed simulator from tens of minutes (even hours) to a few seconds for large-scale problems. During the iterative solution stage of the VoxCap, the compressed circulant tensors are restored/decompressed to their original format *one-by-one* and used in MVMs. Doing so significantly reduces the overall memory cost of the proposed simulator while imposing negligible computational overhead arising from the restoration/decompression operation.

The accuracy and efficiency of the proposed VoxCap simulator are demonstrated (and compared with FastCap [19] when applicable) through its application to the capacitance extraction of various structures including a dielectric-coated perfect electric conductor (PEC) sphere, a parallel interconnect structure, a crossing bus structure, and a parallel meander line structure. The capability of the VoxCap simulator for solving very large-scale problems on a desktop computer is shown by its application to the capacitance extraction of a very large-scale parallel meander line structure discretized by more than one hundred million panels. Moreover, the memory saving achieved by and computational overhead imposed by the Tucker enhancement are extensively quantified for 3D and 2D-like structures. The numerical results show that the proposed VoxCap outperforms the FastCap for many practical scenarios comprising densely packed interconnects. In particular, the VoxCap requires 23x and 47x less memory as well as 11x and 11.9x less CPU time compared to FastCap for the same level of accuracy in the analyses of parallel interconnect structure and parallel meander line structure, respectively. For the analysis of dielectric-coated cube discretized by 400 voxels along $x$-, $y$-, and $z$- directions, the Tucker enhancement reduces the memory requirement of the circulant tensors more than 10,000x while imposing a negligible computational overhead, one-sixth of the CPU time required for one convolution. For the same problem, the Tucker enhancement allows achieving more than 4655x speed-up for obtaining the Toeplitz tensors. Moreover, the scaling of the memory requirement of the Tucker-compressed tensors with respect to increasing computational domain size is found to be sub-linear, while that of the memory requirement of the original circulant and Toeplitz tensors are linear. It should be noted here that a presentation, which sketches the basic principles of the proposed simulator, has been given at an earlier symposium [26]. The detailed information on the simulator is presented for the first time in this paper.

The proposed VoxCap simulator requires $O(N_t \log N_t)$ computational and $O(N_t)$ memory resources, where $N_t$ denotes the number of voxels. Although these cost estimates appear to be not favorable compared to $O(N)$ computational and memory cost estimates of the FastCap, the proposed VoxCap simulator is much faster and less memory demanding compared to the FastCap for many practical scenarios, where $N$ is the number of boundary panels. This is because the multiplicative factors inherent in the cost estimates of the VoxCap are much smaller than those in the cost estimates of the FastCap. To this end, especially when $N_t$ is comparable to $N$ (as in practical interconnect scenarios in Sections III.C, III.D, and III.E), the VoxCap is faster than FastCap. On the other hand, when $N_t$ is much larger than $N$ (as in sphere example in Section III.A), the FastCap is expected to be faster than VoxCap. Note that these facts valid for the FFT-accelerated simulators were also discussed in [20].

## II. FORMULATION

In this section, the SIEs solved by VoxCap, their discretization, and resulting LSE are explained first. Next, the implementation details of the FFTs for the MVMs required during the iterative solution of LSE are provided. Then the proposed preconditioner for reducing the number of iterations during the iterative solution of LSE is described. Finally, the proposed Tucker decomposition/compression scheme is expounded.

*A. SIEs and Their Discretization*

Let $S_c$ and $S_d$ denote PEC and dielectric surfaces with the charge densities $\sigma_c$ and $\sigma_d$, respectively [Fig. 1(a)]. The dielectric surface with outward normal $\hat{\mathbf{n}}_d$ separates the dielectric region with permittivity $\varepsilon_d$ from the background medium with permittivity $\varepsilon_b$. The PEC and dielectric surfaces are assumed to be enclosed by a bounding box [Fig. 1(b)] that consists of $N_t$ voxels with the edge size $\Delta v$, residing on a structured grid; $N_t = N_x \times N_y \times N_z$, where $N_x$, $N_y$, and $N_z$ denote the number of voxels along $x$-, $y$-, and $z$-directions, respectively. In this setting, the capacitance of the structure is computed by solving SIEs [27], which read

$$\Phi(\mathbf{r}) = \int_{S_c} \sigma_c(\mathbf{r}') G(\mathbf{r},\mathbf{r}') dS' + \int_{S_d} \sigma_d(\mathbf{r}') G(\mathbf{r},\mathbf{r}') dS', \quad \mathbf{r} \in S_c \quad (1)$$

$$\varepsilon_d \frac{\partial \Phi^+(\mathbf{r})}{\partial n_d} = \varepsilon_b \frac{\partial \Phi^-(\mathbf{r})}{\partial n_d}, \quad \mathbf{r} \in S_d, \quad (2)$$

where $G(\mathbf{r},\mathbf{r}') = 1/(4\pi\varepsilon_0 |\mathbf{r}-\mathbf{r}'|)$ is the free-space Green's function, $\mathbf{r}$ and $\mathbf{r}'$ denote the observer and source points on the surfaces, respectively, and $\varepsilon_0$ is the permittivity of free space. While $\Phi(\mathbf{r})$ denotes the potential on the surface, $\Phi^+(\mathbf{r})$ and $\Phi^-(\mathbf{r})$ represent the potentials while approaching to the surface from dielectric and background media, respectively. (Note: $\sigma_c$ of the PEC surfaces embedded in a dielectric medium is to be scaled with the permittivity of the dielectric medium, as discussed below.) To solve (1) and (2), $\sigma_c$ and $\sigma_d$ are discretized using the piecewise constant basis functions $w_l$ defined on $N_c$ conductor and $N_d$ dielectric panels as

$$\sigma_c(\mathbf{r}') \cong \sum_{l=1}^{N_c} w_l(\mathbf{r}') \rho_{c,l}, \quad \sigma_d(\mathbf{r}') \cong \sum_{l=N_c+1}^{N_c+N_d} w_l(\mathbf{r}') \rho_{d,l} \quad (3)$$

where $w_l(\mathbf{r}') = 1$ for $\mathbf{r}' \in S_l$; $w_l(\mathbf{r}') = 0$, otherwise. $S_l$ denotes the surface of $l^{\text{th}}$ panel and $\rho_{c,l}$ and $\rho_{d,l}$ are the unknown coefficients of the charge density on the PEC and dielectric panels, respectively. Substituting (3) into (1)-(2), applying Galerkin testing to the resulting equations with $w_k(\mathbf{r})$, $k = 1,\ldots,N$, and evaluating limits for (2) when $l = k$ [28] yields an $N \times N$ LSE as ($N = N_c + N_d$)

$$\begin{bmatrix} \mathbf{V} \\ 0 \end{bmatrix} = \left( \begin{bmatrix} 0 \\ \overline{\mathbf{I}} \end{bmatrix} + \begin{bmatrix} \overline{\mathbf{P}} \\ \overline{\mathbf{E}} \end{bmatrix} \right) \begin{bmatrix} \boldsymbol{\rho}_c \\ \boldsymbol{\rho}_d \end{bmatrix}, \quad (4)$$

where $\boldsymbol{\rho}_c = [\rho_{c,1}, \rho_{c,2}, \ldots, \rho_{c,N_c}]^T$, $\boldsymbol{\rho}_d = [\rho_{d,1}, \rho_{d,2}, \ldots, \rho_{d,N_d}]^T$ are the charge coefficient vectors, $\overline{\mathbf{P}}$ and $\overline{\mathbf{E}}$ matrices with dimensions $N_c \times N$ and $N_d \times N$ relate the potentials and electric fields tested on the panels with the charges, respectively. The entries of $\overline{\mathbf{P}}_{kl}$, $k = 1,\ldots,N_c$, $l = 1,\ldots,N$, are

$$\overline{\mathbf{P}}_{kl} = \int_{S_k} \int_{S_l} w_k(\mathbf{r}) w_l(\mathbf{r}') G(\mathbf{r},\mathbf{r}') dS' dS, \quad (5)$$

and those of $\overline{\mathbf{E}}_{kl}$ and diagonal matrix $\overline{\mathbf{I}}_{kl}$, $k = 1,\ldots,N_d$, $k' = k + N_c$, $l = 1,\ldots,N$, are obtained for non-overlapping and overlapping panels, respectively, via

$$\overline{\mathbf{E}}_{kl} = \frac{\partial}{\partial n_{k'}} \int_{S_{k'}} \int_{S_l} w_{k'}(\mathbf{r}) w_l(\mathbf{r}') G(\mathbf{r},\mathbf{r}') dS' dS, \quad (6)$$

$$\overline{\mathbf{I}}_{kl} = \left( A_{k'}(\varepsilon_d + \varepsilon_b) \right) / \left( 2\varepsilon_0(\varepsilon_d - \varepsilon_b) \right). \quad (7)$$

Here $A_{k'}$ and $\partial/\partial n_{k'}$ represent the area of $S_{k'}$ and partial derivative along the normal to $S_{k'}$, respectively. $\mathbf{V}_k = A_k \Phi_k$, $k = 1,\ldots,N_c$, $\Phi_k$ is the potential applied to the panel on the conductors. For near panel interactions, the integrals in (5) and (6) are evaluated via analytical formulae in Appendix C of [29] and Appendix A, respectively. For far panel interactions, those integrals are evaluated using numerical quadrature and differentiation [28]. (Note: Although the method is explained on a single dielectric medium, it is straightforwardly extended for multi-dielectric media by properly setting the permittivity values in (7).)

To compute the self and mutual capacitances of $m$ conductors, a unit potential is applied to each conductor separately while the potential of the remaining conductors is set to zero, and the LSE in (4) is solved for charge densities. The resulting $m$ number of $\mathbf{V}$ and $\boldsymbol{\rho}_c$ vectors are stored in matrices as $\overline{\mathbf{V}} = [\mathbf{V}^1, \mathbf{V}^2, \ldots, \mathbf{V}^m]$ and $\overline{\boldsymbol{\rho}}_c = [\boldsymbol{\rho}_c^1, \boldsymbol{\rho}_c^2, \ldots, \boldsymbol{\rho}_c^m]$, which are then used to compute $m \times m$ capacitance matrix $\overline{\mathbf{C}}$ as

$$\overline{\mathbf{C}} = \overline{\mathbf{V}}^T \overline{\boldsymbol{\rho}}_c. \quad (8)$$

For the conductor embedded in a dielectric medium, its surface charge is converted to free charge density by multiplying it with $\varepsilon_d / \varepsilon_0$ and used in (8) [28].

The LSE in (4) is iteratively solved for $\boldsymbol{\rho}_c$ and $\boldsymbol{\rho}_d$ via a sequence of MVMs. During the iterative solution, the computational cost of MVMs and the memory requirement of the system matrices scale with $O(N^2)$; the former and the latter are reduced to $O(N_t \log N_t)$ and $O(N_t)$, respectively, via the FFT method explained next. Such computational cost and memory requirements are highly favorable for the capacitance extraction of the densely packed interconnects [20].

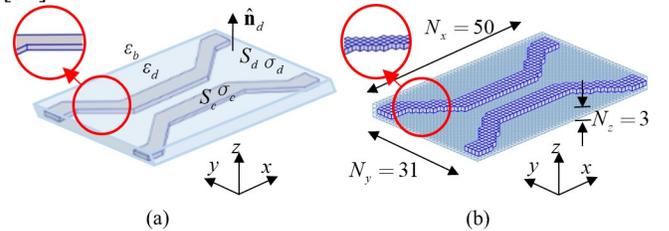

Fig. 1. An example scenario: (a) A structure consisting of interconnects in a dielectric substrate and (b) its discretization via voxels.

*B. FFT Acceleration*

In the FFT acceleration technique, the multiplications of matrices $\overline{\mathbf{P}}$ and $\overline{\mathbf{E}}$ with the vectors $\boldsymbol{\rho}_c$ and $\boldsymbol{\rho}_d$ are considered by grouping the panel interactions with respect to their unit normal as

$$\begin{bmatrix} \mathbf{V}^x \\ 0 \\ \mathbf{V}^y \\ 0 \\ \mathbf{V}^z \\ 0 \end{bmatrix} = \left( \begin{bmatrix} 0 \\ \bar{\mathbf{I}} \\ 0 \\ \bar{\mathbf{I}} \\ 0 \\ \bar{\mathbf{I}} \end{bmatrix} + \begin{bmatrix} \bar{\mathbf{P}}^{x,x} & \bar{\mathbf{P}}^{x,y} & \bar{\mathbf{P}}^{x,z} \\ \bar{\mathbf{E}}^{x,x} & \bar{\mathbf{E}}^{x,y} & \bar{\mathbf{E}}^{x,z} \\ \bar{\mathbf{P}}^{y,x} & \bar{\mathbf{P}}^{y,y} & \bar{\mathbf{P}}^{y,z} \\ \bar{\mathbf{E}}^{y,x} & \bar{\mathbf{E}}^{y,y} & \bar{\mathbf{E}}^{y,z} \\ \bar{\mathbf{P}}^{z,x} & \bar{\mathbf{P}}^{z,y} & \bar{\mathbf{P}}^{z,z} \\ \bar{\mathbf{E}}^{z,x} & \bar{\mathbf{E}}^{z,y} & \bar{\mathbf{E}}^{z,z} \end{bmatrix} \right) \begin{bmatrix} \boldsymbol{\rho}^x \\ \boldsymbol{\rho}^y \\ \boldsymbol{\rho}^z \end{bmatrix}, \quad (9)$$

where $\boldsymbol{\rho}^x = [\boldsymbol{\rho}_c^x; \boldsymbol{\rho}_d^x]$, $\boldsymbol{\rho}^y = [\boldsymbol{\rho}_c^y; \boldsymbol{\rho}_d^y]$, and $\boldsymbol{\rho}^z = [\boldsymbol{\rho}_c^z; \boldsymbol{\rho}_d^z]$ are the vectors of the charge coefficients of the panels with unit normal pointing along $x$-, $y$-, and $z$- directions. Here the block matrix $\bar{\mathbf{P}}^{y,x}$, for example, stores the potentials generated by the charges on the panels with unit normal pointing along $x$-direction and tested on panels with unit normal pointing along the $y$-direction. The full block matrices are multiplied with charge coefficient vectors as

$$\mathbf{C}^\alpha = \sum_\beta \bar{\mathbf{P}}^{\alpha,\beta} \boldsymbol{\rho}^\beta, \quad \mathbf{D}^\alpha = \sum_\beta \bar{\mathbf{E}}^{\alpha,\beta} \boldsymbol{\rho}^\beta \quad (10)$$

where $\alpha, \beta \in \{x, y, z\}$. Note that the multiplications of $\boldsymbol{\rho}^\beta$ with the diagonal matrix $\bar{\mathbf{I}}$ in (9) are computationally cheap and performed traditionally. However, the multiplications of $\boldsymbol{\rho}^\beta$ with the full block matrices $\bar{\mathbf{P}}^{\alpha,\beta}$ and $\bar{\mathbf{E}}^{\alpha,\beta}$ are computationally expensive and thereby accelerated via FFT technique. In this technique, the multiplications are performed by taking into account the panels of $N_t$ voxels instead of $N$ boundary panels on the PEC and dielectric surfaces. To do that, first, $(N_x+1) \times N_y \times N_z$, $N_x \times (N_y+1) \times N_z$, and $N_x \times N_y \times (N_z+1)$ numbers of voxel panels with unit normal pointing along $x$-, $y$-, and $z$-directions are identified, respectively. These voxel panels lie on structured grids and thereby their interactions are characterized by block Toeplitz tensors. The block Toeplitz tensors $\mathcal{A}^{\alpha,\beta}$ and $\mathcal{B}^{\alpha,\beta}$ are used to obtain block circulant tensors $\mathcal{P}^{\alpha,\beta}$ and $\mathcal{E}^{\alpha,\beta}$ corresponding to blocks $\bar{\mathbf{P}}^{\alpha,\beta}$ and $\bar{\mathbf{E}}^{\alpha,\beta}$, $\alpha, \beta \in \{x, y, z\}$. The procedure to obtain the block Toeplitz tensors and the circulant tensors are explained in Appendix B. The block circulant tensors are Fourier transformed and stored as $\tilde{\mathcal{P}}^{\alpha,\beta}$ and $\tilde{\mathcal{E}}^{\alpha,\beta}$ during the setup stage of the simulator. During the iterative solution stage, they are used to perform the MVMs corresponding to each block as

$$\mathcal{C}^\alpha = IFFT\left\{\sum_\beta \tilde{\mathcal{P}}^{\alpha,\beta} * \tilde{\mathcal{Q}}^\beta\right\}, \mathcal{D}^\alpha = IFFT\left\{\sum_\beta \tilde{\mathcal{E}}^{\alpha,\beta} * \tilde{\mathcal{Q}}^\beta\right\}, (11)$$

where * denotes the tensor-tensor multiplication, $IFFT\{\cdot\}$ is the inverse FFT operator, and $\tilde{\mathcal{Q}}^\beta = FFT\{\mathcal{Q}^\beta\}$. The tensor $\mathcal{Q}^\beta$ with the dimensions $2(N_x+1) \times 2(N_y+1) \times 2(N_z+1)$ is filled by the samples from $\boldsymbol{\rho}^\beta$ and zeros [30]. The results of MVMs, $\mathbf{C}^\alpha$ and $\mathbf{D}^\alpha$, are obtained from the entries of $\mathcal{C}^\alpha$ and $\mathcal{D}^\alpha$.

A couple of notes regarding the computation of circulant tensors and their deployment in FFT operations are in order:
1) Although the dimensions of each circulant tensor vary due to the different numbers of voxel panels aligned along $x$-, $y$-, and $z$-directions, the dimensions of all circulant tensors are enlarged to $2(N_x+1) \times 2(N_y+1) \times 2(N_z+1)$ by carefully padding zeros, as explained in Appendix B. Such zero-padding enables to reduce the number of FFT operations from nine to three while computing $\tilde{\mathcal{Q}}^\beta$ and the number of IFFT operations from nine to three while computing each of $\mathcal{C}^\alpha$ and $\mathcal{D}^\alpha$, $\alpha \in \{x, y, z\}$.
2) By exploiting the symmetry and invoking the properties of Fourier transform [25], some of the FFT'ed circulant tensors are obtained from others via complex conjugation as

$$\tilde{\mathcal{P}}^{y,x} = conj\{\tilde{\mathcal{P}}^{x,y}\}, \tilde{\mathcal{P}}^{z,x} = conj\{\tilde{\mathcal{P}}^{x,z}\}, \tilde{\mathcal{P}}^{z,y} = conj\{\tilde{\mathcal{P}}^{y,z}\}. \quad (12)$$

During tensor-tensor multiplication in (11), the conjugation is performed on the fly and thereby the memory is not required to store $\tilde{\mathcal{P}}^{y,x}$, $\tilde{\mathcal{P}}^{z,x}$, and $\tilde{\mathcal{P}}^{z,y}$.
3) The directions of the panels' unit normal should be carefully taken into account while obtaining the entries of $\mathbf{D}^\alpha$ from $\mathcal{D}^\alpha$. The entries of $\mathbf{D}^\alpha$ for boundary panels with unit normal pointing along positive $x$-, $y$-, and $z$- directions are directly retrieved from the entries of $\mathcal{D}^\alpha$, whereas those for boundary panels with unit normal pointing along negative $x$-, $y$-, and $z$-directions are obtained from the entries of $\mathcal{D}^\alpha$ after flipping the signs of entries.

*C. Tucker Enhancement*

The proposed VoxCap simulator uses Tucker decompositions to compress the Toeplitz tensors $\mathcal{A}^{\alpha,\beta}$ and $\mathcal{B}^{\alpha,\beta}$, $\alpha, \beta \in \{x, y, z\}$, *once and for all* during the installation stage. During the setup stage of the simulator's execution, the compressed Toeplitz tensors are read from the hard disk and used to obtain the circulant tensors. Doing so allows reducing the setup stage time of the simulator from tens of minutes (and even hours) to seconds, as shown in Table I of Section III.B.3. To do that, during the installation stage of the simulator, the Toeplitz tensors $\mathcal{A}^{\alpha,\beta}$ and $\mathcal{B}^{\alpha,\beta}$ are computed for a large computational domain (say $N_x = N_y = N_z = 1,000$) by setting $\Delta v = 1$ m, compressed by Tucker decompositions, and stored on the hard disk. During the setup stage of each execution, the compressed tensors requiring megabytes of memory are read from the hard disk and restored to their original format. The restored Toeplitz tensors are resized with respect to the dimensions of the computational domain required for the structure being analyzed. Next, the restored Toeplitz tensors are multiplied with scaling factors related to the voxel size. These scaling factors, derived in Appendix C, are $(\Delta v)^3$ and $(\Delta v)^2$ for $\mathcal{A}^{\alpha,\beta}$ and $\mathcal{B}^{\alpha,\beta}$, respectively. Finally, the restored Toeplitz tensors are used in the embedding procedure explained in Appendix B to obtain circulant tensors.

In addition, the proposed VoxCap simulator leverages Tucker decompositions to reduce the memory requirement of the FFT'ed circulant tensors $\tilde{\mathcal{P}}^{\alpha,\beta}$ and $\tilde{\mathcal{E}}^{\alpha,\beta}$, $\alpha, \beta \in \{x, y, z\}$. Doing so allows reducing the memory footprint of the simulator around a factor of three and half for the capacitance extraction of dielectric coated PEC structures, as shown in numerical example in Section III.A. To do that, during the setup stage of the simulator, all $\tilde{\mathcal{P}}^{\alpha,\beta}$ and $\tilde{\mathcal{E}}^{\alpha,\beta}$ $\alpha,\beta \in \{x,y,z\}$ are compressed by Tucker decomposition. During the iterative solution stage, each compressed tensor is restored/decompressed to its original format *one-by-one*. The computational penalty associated with the restoration operation is negligible compared to one convolution operation perform during MVMs, as shown in the numerical results section. It should be noted that the abovementioned Tucker enhancement strategies have recently been applied to the data structure involving the magneto-quasi-static kernels in VoxHenry as well [31].

The Toeplitz and FFT'ed circulant tensors are compressed by Tucker decompositions as [24], [32]-[34]

$$\mathcal{X} \approx \mathcal{S} \times_1 \overline{\mathbf{U}}^1 \times_2 \overline{\mathbf{U}}^2 \times_3 \overline{\mathbf{U}}^3, \quad (13)$$

where the tensor $\mathcal{X}$ with dimensions $D_1 \times D_2 \times D_3$ represents $\mathcal{A}^{\alpha,\beta}$, $\mathcal{B}^{\alpha,\beta}$, $\tilde{\mathcal{P}}^{\alpha,\beta}$, or $\tilde{\mathcal{E}}^{\alpha,\beta}$, $\alpha,\beta \in \{x,y,z\}$, $\mathcal{S}$ is the core tensor with dimensions $r_1 \times r_2 \times r_3$, $\overline{\mathbf{U}}^i$, $i=1,\ldots,3$, represents the factor matrices with dimensions $D_i \times r_i$, and $r_i$ is the multilinear rank pertinent to $i^{\text{th}}$ dimension. A tensor is Tucker compressible in case $(r_1 r_2 r_3) + \sum_{i=1}^3 D_i r_i \ll D_1 D_2 D_3$. The symbol $\times_i$, $i=1,\ldots,3$, stands for $i-$ mode matrix-tensor multiplication. The procedure to obtain the $\mathcal{S}$ and $\overline{\mathbf{U}}^i$ for given tolerance, *tol*, is provided in [32], [33]. Typically, *tol* is set to $10^{-8}$ in this study to achieve high compressibility without sacrificing from the accuracy, unless stated otherwise.

### D. Block-Diagonal-Diagonal Preconditioner

The proposed VoxCap simulator uses a block-diagonal-diagonal preconditioner to ensure the rapid convergence of the iterative solution of (4). To this end, during each MVM, the block diagonal preconditioner $\overline{\mathbf{R}}_{BD}$ and diagonal preconditioner $\overline{\mathbf{R}}_D$ are applied to (4) as

$$\begin{bmatrix} \overline{\mathbf{R}}_{BD} \\ \overline{\mathbf{R}}_D \end{bmatrix} \begin{bmatrix} \mathbf{V} \\ 0 \end{bmatrix} = \begin{bmatrix} \overline{\mathbf{R}}_{BD} \\ \overline{\mathbf{R}}_D \end{bmatrix} \left( \begin{bmatrix} 0 \\ \overline{\mathbf{I}} \end{bmatrix} + \begin{bmatrix} \overline{\mathbf{P}} \\ \overline{\mathbf{E}} \end{bmatrix} \right) \begin{bmatrix} \boldsymbol{\rho}_c \\ \boldsymbol{\rho}_d \end{bmatrix} \quad (14)$$

The entries of the diagonal matrix $\overline{\mathbf{R}}_D$ are directly obtained from the inverse of the diagonal matrix $\overline{\mathbf{I}}_{kl}$. The blocks of block-diagonal matrix $\overline{\mathbf{R}}_{BD}$ are formed by (i) splitting the bounding box enclosing the structure into small boxes, (ii) computing the interactions between basis functions in each small box separately, and (iii) assigning the inverses of the submatrices storing the interactions as blocks of $\overline{\mathbf{R}}_{BD}$. Each small box comprises of $N_{vx}$, $N_{vy}$, and $N_{vz}$ voxels along *x*-, *y*-, and *z*-directions. For an example PEC interconnect scenario, this partitioning and the basis functions/panels used to form the blocks are shown in Fig.2 ($N_{vx} = N_{vy} = 16$ and $N_{vz} = 1$). Two important points regarding to the construction and storage of the blocks are in order: (i) The interactions between basis functions in each small box are directly obtained from one-time generated circulant tensors $\mathcal{P}^{\alpha,\beta}$ and $\mathcal{E}^{\alpha,\beta}$, $\alpha,\beta \in \{x,y,z\}$ for a small box consisting of $N_{vx}$, $N_{vy}$, and $N_{vz}$ voxels. This yields significant computational saving while constructing the preconditioner. Note that the circulant tensors store all possible basis function interactions for a small box. By using these tensors, the computational time required to compute (5) for similar basis function pairs in blocks are avoided. (ii) Many blocks in $\overline{\mathbf{R}}_{BD}$ are the replicates of each other for a voxelized structure. To this end, only unique blocks are stored and the memory requirement of the preconditioner is minimized.

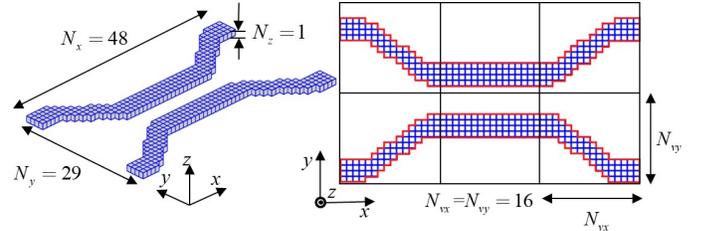

Fig. 2. The partitioning of a voxelized structures (left) via three and two small boxes along *x*- and *y*-directions (right). Each block of $\overline{\mathbf{R}}_{BD}$ is formed by the interactions between boundary panels (indicated by red color) within each small box; the boundary panels with unit normal pointing along *z*-direction are excluded from the right figure for illustration purposes.

### III. NUMERICAL RESULTS

In this section, several numerical examples that show the applicability, memory and CPU efficiency, and accuracy of the proposed VoxCap simulator are presented. In the following analysis, when applicable, the capacitance values obtained by VoxCap and FastCap are compared with each other or those obtained by an analytical formula. The discrepancy between the results is quantified through the relative error defined as $err = |(F - \tilde{F})/\tilde{F}|$, where $F$ and $\tilde{F}$ denote the capacitance values obtained by VoxCap/FastCap and analytical formula, respectively. Furthermore, the charge distributions on the structures are plotted in the logarithmic (dB) scale after all charge values are normalized by the maximum charge value on the structure and the logarithm of normalized values are multiplied by twenty. The proposed VoxCap simulator was implemented in Matlab, while the FastCap simulator executes a C code. The linear system of equations in (4) is iteratively solved by generalized minimal residual method (GMRES) with a restart every 35 iterations until the relative residual error (RRE) reaches to $10^{-4}$, unless stated otherwise. Likewise, FastCap also uses GMRES with the same restart to reach to the same RRE; it uses an overlapped diagonal preconditioner to reduce the number of iterations [28]. All simulations are executed on an Intel Xeon Gold 6412 CPU with 384 GB RAM.

### A. The Dielectric-Coated PEC Sphere

First, the proposed VoxCap simulator and FastCap simulator (with $4^{\text{th}}$ order multipole expansion) are used to obtain the self-capacitance of a 0.25 m - radius PEC sphere coated by a dielectric shell with relative permittivity $\varepsilon_r$ and radius of 0.5 m [Fig. 3(a)]. The structure is discretized by

voxels of size $\Delta v = \{0.05, 0.025, 0.02, 0.01\}$ m, which gives rise to $N_t = \{8,000, 64,000, 125,000, 1,000,000\}$ and $N = \{2,376, 9,480, 14,760, 59,016\}$, respectively. For $\varepsilon_r = 2$ and all voxel sizes, the self-capacitance values computed by the VoxCap and FastCap simulators are compared with those obtained by the analytical formula, which reads as $C = 4\pi\varepsilon_0 \varepsilon_r r_d r_c / ((r_d - r_c) + (\varepsilon_r r_c))$, where $r_c$ and $r_d$ are the radii of PEC sphere and dielectric shell, respectively; the relative difference between results $err$ is plotted [Fig. 3(a)]. Apparently, while $err$ decreases with increasing $1/\Delta v$, the accuracy achieved by both VoxCap and FastCap simulators is nearly the same. Their accuracy stagnates at the same level due to staircase approximation to the spherical shape, which clearly appears in the normalized charge distributions plots for the structure discretized with $\Delta v = \{0.025, 0.01\}$ m [Fig. 3(b)]. For the simulations of the structures discretized with $\Delta v = \{0.05, 0.025, 0.02, 0.01\}$ m, the proposed VoxCap and FastCap simulators required CPU time of $\{1.54, 10.30, 11.75, 155.52\}$ s and $\{0.48, 2.02, 3.51, 14.26\}$ s, iterations of $\{7, 7, 10, 10\}$ and $\{7, 7, 7, 9\}$ during the iterative solution, and the memory of $\{8, 52, 97, 681\}$ MB and $\{135, 612, 992, 3009\}$ MB, respectively. Note that the Tucker enhancement reduces the memory requirement from 2,440 MB to 681 MB (more than a factor of three and half) for the analysis of the structure discretized with $\Delta v = 0.01$ m while imposing a computational penalty of 33%. It is expected that the VoxCap is much more memory efficient compared to FastCap while the FastCap is faster than VoxCap for this validation example with well-separated panels and high $N_t / N$ ratio. It is shown below in the examples with densely-packed interconnects and low $N_t / N$ ratios that the VoxCap is indeed much faster than FastCap for practical scenarios.

Next, $\varepsilon_r$ is swept from 2 to $2 \times 10^7$ with one-decade increment at each simulation of the structure discretized by $\Delta v = 0.05$ m [Fig. 3(c)]. For each simulation, an RRE of $10^{-8}$ is achieved during the iterative solution of (4) to avoid any inaccuracy causing from the iterative solution. In case such RRE is not achieved, accurate results can not be obtained for large $\varepsilon_r$ values. Clearly, the proposed VoxCap produces accurate results for all $\varepsilon_r$ values, while FastCap yields inaccurate results with increasing permittivity, as indicated in [35]. Finally, the effectiveness of the proposed preconditioner is examined for the structure discretized by $\Delta v = 0.01$ m and $\varepsilon_r$ set to 2. For this case, the proposed VoxCap simulator iteratively solves (4) without using a preconditioner and with using the proposed block-diagonal-diagonal preconditioner, a diagonal preconditioner, and block-diagonal preconditioner. The blocks in proposed and block-diagonal preconditioners are formed by setting $N_{vx} = N_{vy} = N_{vz} = 10$. For all these cases,

RRE at each iteration is plotted [Fig. 3(d)]. The number of iterations required to reach $10^{-8}$ is 22, 27, 41 and 142 when the proposed preconditioner, the block diagonal preconditioner, the diagonal preconditioner, and no preconditioner are applied, respectively. Needless to say, the proposed preconditioner outperforms all other preconditioners and yields the fastest convergence. Furthermore, it requires 17.04 MB memory, a quarter of 70.26 MB memory requirement of the conventional block-diagonal preconditioner.

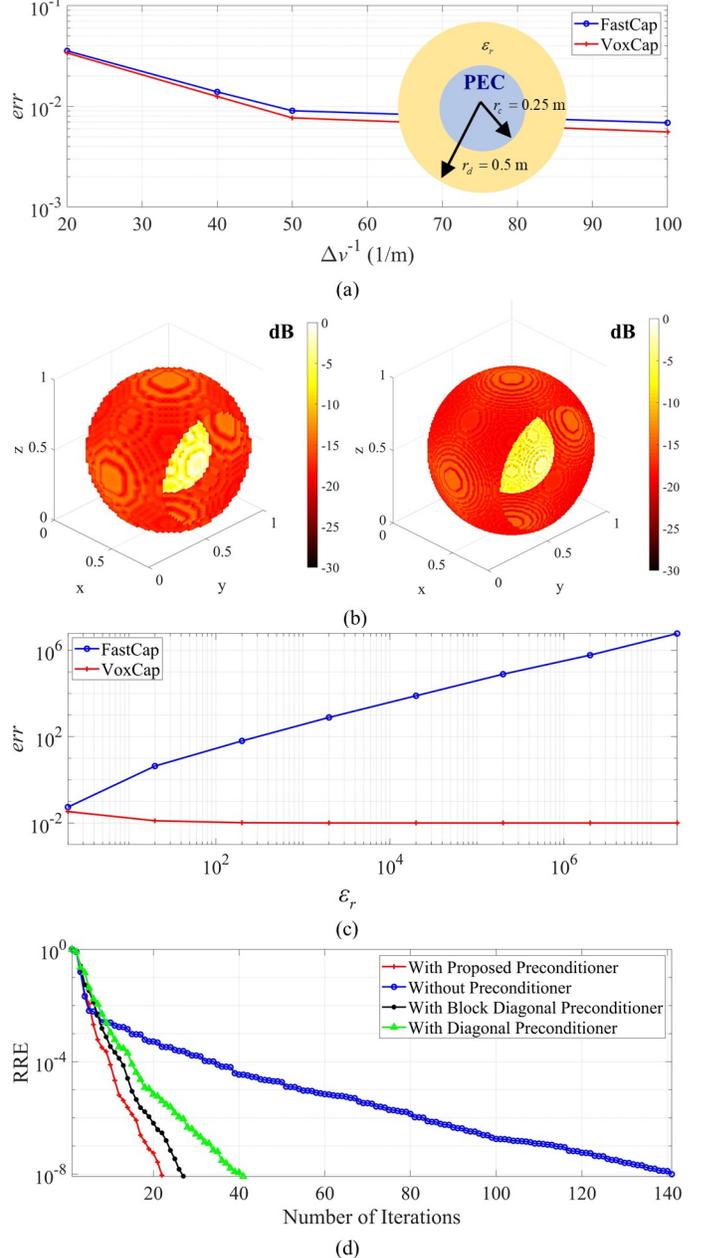

Fig. 3. Dielectric-coated PEC sphere example: (a) The $err$ in the self-capacitance computed by VoxCap and FastCap with decreasing voxel size (for $\varepsilon_r = 2$). (b) The normalized charge distribution on the structure when $\Delta v = 0.025$ m (left) and $\Delta v = 0.01$ m (right). (Note: The panels with x-coordinate greater than 0.85 m are removed for illustration purposes.) (c) The $err$ in self-capacitance obtained by VoxCap and FastCap with increasing $\varepsilon_r$ (for $\Delta v = 0.05$ m). (d) The RRE history of the LSE's iterative solution when the proposed preconditioner, no preconditioner, and conventional preconditioners are applied.

*B. Numerical Tests on Tucker Enhancement*

In this part, the memory saving achieved by and computational overhead introduced by Tucker decomposition are investigated while compressing and restoring the circulant tensors. In addition, the computational saving achieved by obtaining the Toeplitz tensors from their compressed representations is demonstrated. To this end, the memory saving is quantified via compression ratio (CR), which is the ratio of memory requirement of original tensors to the memory requirement of the compressed tensor. Furthermore, the computational overhead (CO) is quantified by taking the ratio of the time for restoring the tensor to the time for performing one convolution with that tensor.

*1) A dielectric coated PEC plate:* First, a dielectric coated PEC plate with varying width and length is considered to test the performance of Tucker compression/decompression for a 2D-like structure [Fig. 4 (a)]. The structure fully enclosed by the computational domain has the length and width varying from $100\,\mu m$ to $1500\,\mu m$ while $\Delta v = 1\,\mu m$. For the structure with varying dimensions, Tucker decomposition is used to compress the circulant tensors; the CR achieved by and CO imposed by the Tucker decomposition are plotted in Figs. 4(a) and (b). Clearly, CR increases with increasing tolerance and the structure size (or number of voxels $N_t$). Tucker enhancement achieves more than 700x reduction (for $tol = 10^{-4}$) in the memory of circulant tensors for the largest structure. Moreover, CO decreases with increasing computational domain size and reaches to 0.05 for the largest structure size, which shows that the computational penalty associated with the tensor restoration operation is negligible. This negligible penalty is due to the fact the multilinear ranks of the compressed tensors remain nearly constant with increasing computational domain size. Fig. 4(c) demonstrates the change of the maximum multilinear rank of $\tilde{\mathcal{P}}^{x,x}$ (as an example) with increasing $N_t$ for different *tol* values. Clearly, the maximum rank is a nearly constant function of $N_t$ and changes between 10 and 30. Fig. 4(d) shows the memory requirement of the original tensors as well as the compressed tensors with increasing structure size. Needless to say, the memory of original tensors scales with $O(N_t)$ while the memory of compressed tensors scales with $O(N_t^{0.44} \log(N_t))$.

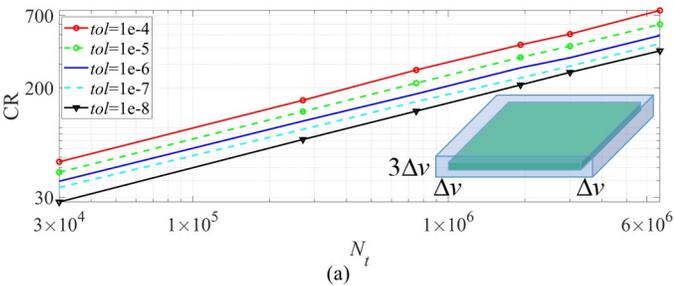

(a)

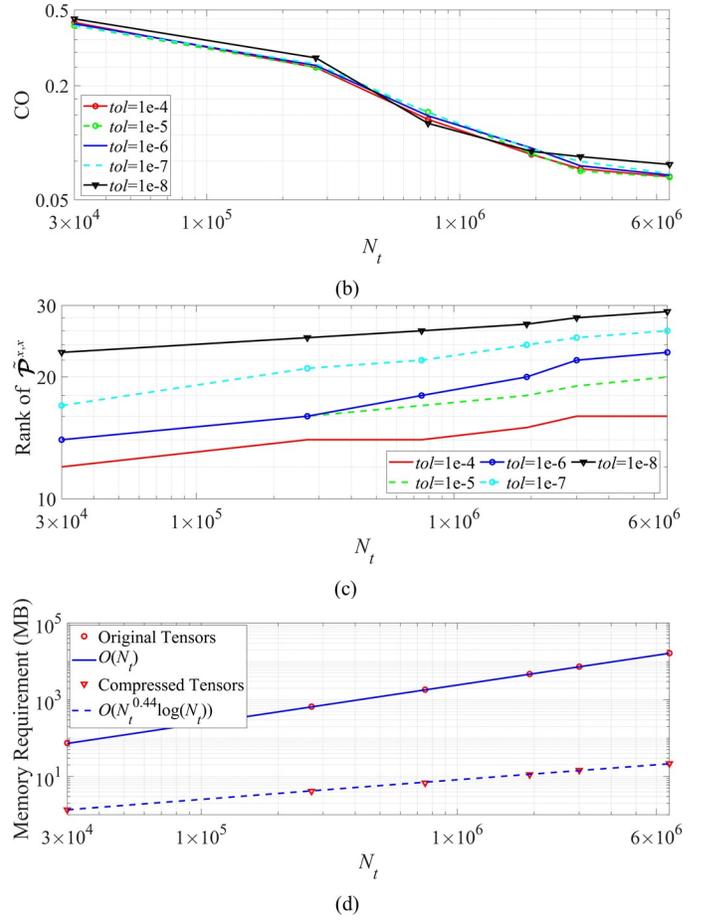

(b)

(c)

(d)

Fig 4. The performance of Tucker compression/decompression on a 2D-like dielectric-coated PEC plate. (a) CR achieved by and (b) CO introduced by the Tucker enhancement with increasing $N_t$. (c) The maximum multilinear rank of $\tilde{\mathcal{P}}^{x,x}$ and (d) the memory scaling of the original and compressed FFT'ed circulant tensors with increasing $N_t$.

*2) A dielectric coated cube:* Next, a dielectric coated PEC cube with varying edge length is considered for testing the performance of Tucker compression/decompression for a 3D structure [Fig. 5(a)]. The cube fully enclosed by the computational domain has the edge length changed from $63\,\mu m$ to $400\,\mu m$ while $\Delta v = 1\,\mu m$. For the structure with varying edge length, the circulant tensors are compressed by Tucker decompositions. The CR achieved by and CO introduced by Tucker compression and decompression are plotted in Figs. 5(a) and (b), respectively. Again, the CR increases with increasing tolerance and structure size (or number of voxels $N_t$). Tucker enhancement achieves more than 10000x (four orders of magnitude) reduction (for $tol = 10^{-4}$) in the memory of circulant tensors for the largest structure. Compared to dielectric coated PEC plate case, this dramatic reduction is expected as the tensor methods yield much better compression for the tensors with many elements along all three dimensions compared to the tensors with many elements along two dimensions and less elements in the remaining dimension, as in dielectric coated PEC plate case. Moreover, CO decreases with increasing structure size and reaches to 0.154 for the largest structure size for different *tol* values. This again shows that the computational penalty associated with the tensor restoration operation is negligible. Fig. 5(c) shows the change of the maximum multilinear rank of $\tilde{\mathcal{P}}^{x,x}$ with increasing $N_t$ for different *tol* values. Clearly, the maximum rank negligibly increases in the range from 10 to 20. Fig. 5(d) demonstrates the memory requirement of the original tensors as well as the compressed tensors with increasing structure size. While the memory of original tensors scales with $O(N_t)$, the memory of compressed tensors scales with $O(N_t^{0.26} \log(N_t))$, which is better than $O(N_t^{0.44} \log(N_t))$ obtained for the dielectric-coated PEC plate.

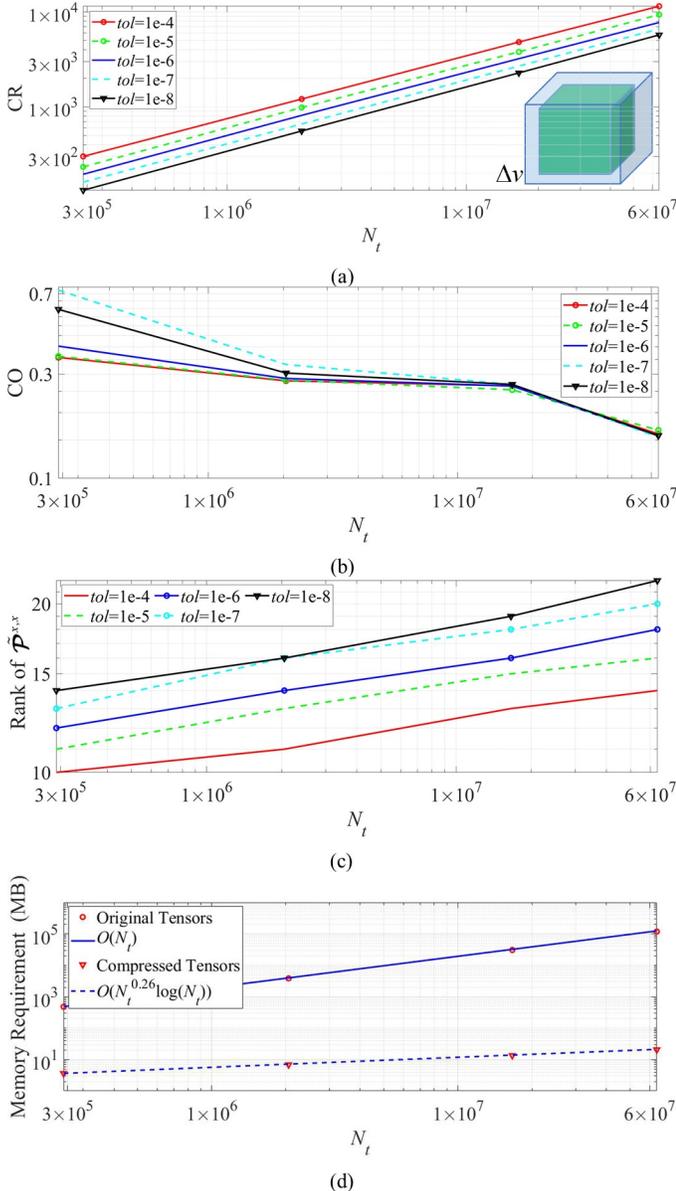

Fig. 5. The performance of Tucker compression/decompression on a 3D dielectric-coated PEC cube. (a) CR achieved by and (b) CO introduced by the Tucker enhancement with increasing $N_t$. (c) The maximum multilinear rank of $\tilde{\mathcal{P}}^{x,x}$ and (d) the memory scaling of the original and compressed FFT'ed circulant tensors with increasing $N_t$.

*3) Performance on Toeplitz Tensor:* Next, the computational time saving achieved by obtaining Toeplitz tensors from their Tucker-compressed versions is quantified. To this end, the dielectric coated cube in the previous analysis is considered. Its edge length is changed from 100 μm to 400 μm with increment of 100 μm while $\Delta v = 1$ μm. For the analysis of cube with different edge lengths, the memory requirement of the original Toeplitz tensors and Tucker-compressed Toeplitz tensors are tabulated in Table I. Furthermore, the CPU time required to generate original Toeplitz tensor and the CPU time required to obtain the Toeplitz tensor from Tucker-compressed Toeplitz tensors are provided. Apparently, the compressed tensors stored on hard disk require a few MBs memory for $tol = 10^{-8}$ while the original ones require around 1 GB memory for the largest structure. While the CPU time to obtain the Toeplitz tensors is around 1.3 hours for the largest structure, the total CPU time to obtain the Toeplitz tensors from compressed tensors (including the CPU time for reading from harddisk and restoration from the compressed tensors) is 1.04 second for the same structure. By obtaining the Toeplitz tensors from Tucker compressed ones, 4655x speed-up is achieved for the largest structure.

TABLE I
MEMORY AND CPU TIME REQUIREMENT FOR OBTAINING TOEPLITZ TENSORS BY COMPUTATION AND FROM TUCKER-COMPRESSED TENSORS

| Edge length of cube (μm) | Memory of original Toeplitz tensors (MB) | Memory for compressed tensors (MB) | CPU time to generate original Toeplitz tensors (s) | CPU time to obtain Toeplitz tensors from compressed tensors (s) |
|---|---|---|---|---|
| 100 | 15.72 | 1.57 | 85.27 | 0.21 |
| 200 | 123.91 | 2.76 | 616.39 | 0.58 |
| 300 | 416.12 | 3.91 | 1858.59 | 0.822 |
| 400 | 983.91 | 5.10 | 4841.59 | 1.04 |

### C. Parallel Interconnects

In this practical example, a five by twelve parallel interconnect array embedded in a dielectric substrate ($\varepsilon_r = 7$) with 0.02 mm spacing from the surfaces of the substrate is considered [Fig. 6(a)]. The width, length, and height of each interconnect are 0.08, 0.8, and 0.02 mm, respectively, while the center-to-center distance between interconnects is 0.1 mm. The discretization of the structure with $\Delta v = 0.005$ mm yields $N_t = 1,780,800$ and $N = 481,024$ (note that $N_t / N = 3.7$). The VoxCap with $N_{vx} = N_{vy} = N_{vz} = 10$ and FastCap with 2nd, 4th, and 6th order multipole expansions are used to obtain the capacitance matrix for the structure. Figs. 6(a)-(c) compare the values in the first column of the capacitance matrix obtained by the VoxCap and FastCap with 2nd, 4th, and 6th order multipole expansions. Clearly, only the FastCap with 6th order multipole expansion provides accurate results while the results obtained by the FastCap with 2nd and 4th order multipole expansions are highly inaccurate. The maximum relative differences between results of VoxCap and FastCap with 2nd, 4th, and 6th order multipole expansions are 15.96, 7.54, and 0.15, respectively. For this example, VoxCap and FastCap with 2nd, 4th, and 6th order multipole expansions required 2413, 2843, 2064, and 2063 iterations during the iterative solution, respectively. Fig. 6(c) also demonstrates the normalized charge distribution obtained by VoxCap when the structure is excited from the 45th conductor (the conductors are numbered from left to right starting from left bottom corner). Table II compares the memory and computational time requirements of VoxCap simulator and FastCap simulator with 2nd, 4th, and 6th order multipole expansions. For the same level of accuracy, the VoxCap with Tucker enhancement required 23x less memory and 11x less CPU time compared to the FastCap with 6th order multipole expansion. In this example, Tucker enhancement achieves more than 502x reduction in the memory of circulant tensors. Furthermore, it reduces the memory requirement of the simulator from 5.54 GB to 2.48 GB by a factor 2.23 while increasing the CPU time requirement by 10%.

### D. Crossing Buses

In this example, a crossing bus structure with six layers of dielectrics [36] is considered [Fig. 7(a)]. In this structure, 2nd, 4th, and 6th layers consist of a 40 nm-thick dielectric with $\varepsilon_r = 5.0$, while the 1st, 3rd, and 5th layers, each of which houses 15 buses, are filled with 220 nm-thick dielectric with $\varepsilon_r = 2.6$. Moreover, the buses in the 3rd layer are coated by a 10 nm-thick dielectric with $\varepsilon_r = 3.7$. The width, height, and length of each bus are 70 nm, 140 nm, and 2,030 nm, respectively, while their center-to-center distance is 140 nm. The distance between the leftmost/rightmost buses and the surfaces of the dielectrics is 70 nm. The structure is discretized by $\Delta v = 10$ nm. This discretization yields $N_t = 3,672,942$ and $N = 849,847$ ($N_t / N = 4.32$). VoxCap with $N_{vx} = N_{vy} = N_{vz} = 10$ and FastCap with 2nd and 4th order multipole expansions are used to obtain the capacitance matrix for this structure. Figs. 7(a)-(b) compare the values in the first column of the capacitance matrix obtained by VoxCap and FastCap with 2nd and 4th order multipole expansions. Fig. 7(b) also shows the normalized charge distribution obtained by VoxCap when the structure is excited from the foremost conductor (with the center coordinates of (1085,105,590) nm) in the 5th layer. Clearly, FastCap with 4th order multipole expansion provides accurate results while the results obtained by FastCap with 2nd order multipole expansions are inaccurate. The maximum relative differences between results of VoxCap and FastCap with 2nd and 4th order multipole expansions are 1.0069 and 0.091, respectively. For this example, VoxCap and FastCap with 2nd and 4th order multipole expansions required 1546, 2512, and 1637 iterations, respectively. Table III compares the memory and computational time requirements of VoxCap simulator and FastCap simulator with 2nd and 4th order multipole expansions. For the same level of accuracy, the VoxCap with Tucker enhancement required 23x less memory and 10x less CPU time compared to the FastCap with 4th order multipole expansion. In this example, Tucker enhancement achieves more than 801x reduction in the memory of circulant tensors. Furthermore, it reduces the memory requirement of the simulator from 10.37 GB to 3.75 GB by a factor 2.76 while increasing the CPU time requirement by 29.8%. Although VoxCap permits efficient and accurate analysis of the structures consisting of multiple thin/thick layers and conductors, its performance could decline for the analysis of the structures with very thin dielectrics/conductors and multiscale features due to increasing $N_t / N$ ratio, as do all voxel-based techniques.

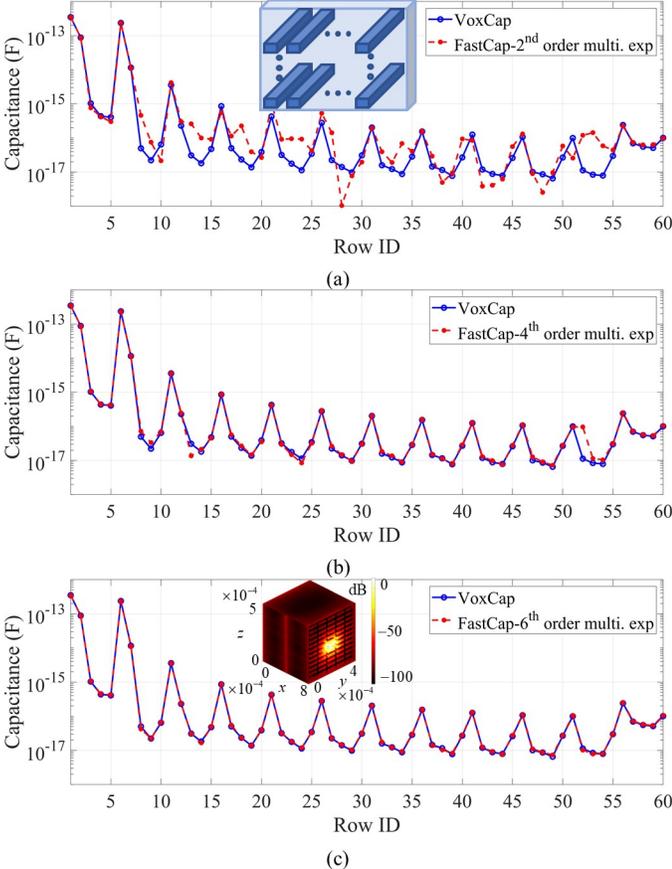

Fig. 6. Parallel interconnect example. The comparison of the values in the first column of the capacitance matrix obtained by the VoxCap and the FastCap with (a) 2nd order and (b) 4th order multipole expansions, as well as (c) 6th order multipole expansions and normalized charge distribution on the structure. (Note: For the illustration purposes, the panels on the dielectric substrate with x-coordinates larger than 0.83 mm were removed.)

TABLE II
MEMORY AND CPU TIME REQUIREMENTS OF VOXCAP AND FASTCAP SIMULATORS FOR PARALLEL INTERCONNECT EXAMPLE

|  | Memory requirement (MB) | CPU time requirement (s) |
| --- | --- | --- |
| VoxCap (w/out Tucker) | 5,536 | 4,613 |
| VoxCap (w/ Tucker) | 2,473 | 5,080 |
| FastCap w/ 2nd order expansion | 17,200 | 14,647 |
| FastCap w/ 4th order expansion | 33,400 | 29,352 |
| FastCap w/ 6th order expansion | 56,900 | 57,644 |

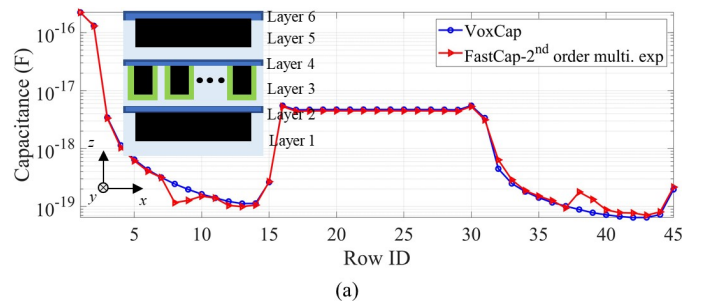

(a)

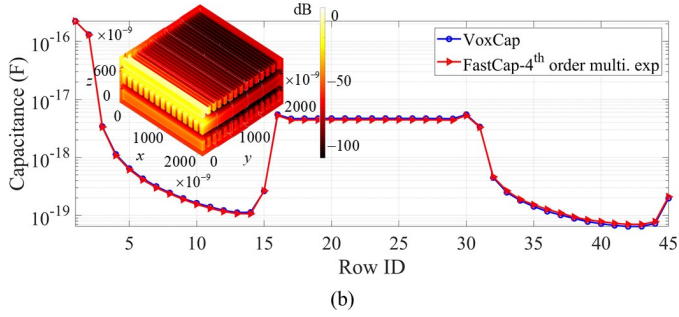

(b)

Fig. 7. Crossing bus example. The comparison of the values in the first column of the capacitance matrix obtained by the VoxCap and the FastCap with (a) $2^{nd}$ order and (b) $4^{th}$ order multipole expansions and the normalized charge distribution on the structure. (Note: For illustration purposes, the panels on the dielectric substrate with $x$-coordinates larger than 2105 nm, $y$-coordinates less than 65 nm, $z$-coordinates larger than 750 nm were removed.).

TABLE III
MEMORY AND CPU TIME REQUIREMENTS OF VOXCAP AND FASTCAP SIMULATORS FOR CROSSING BUS EXAMPLE

| | Memory requirement (MB) | CPU time requirement (s) |
|---|---|---|
| VoxCap (w/out Tucker) | 10,372 | 5,729 |
| VoxCap (w/ Tucker) | 3,755 | 7,437 |
| FastCap w/ $2^{nd}$ order expansion | 36,500 | 43,396 |
| FastCap w/ $4^{th}$ order expansion | 86,900 | 77,021 |

*E. Parallel Meander Lines*

Next, the parallel meander line structure is considered to check the accuracy and capabilities of the proposed VoxCap simulator [Fig. 8]. The structure is formed by the interconnects with the width $w_b$, length $l_b$, height $h_b$, and spacing $d$. There exist $n$ number of interconnects in each of $n$ number of layers. For the first analysis, after setting $w_b = h_b = 0.5$ mm, $d = 0.1$ mm, $l_b = 10$ mm, and $n = 20$, the structure is discretized by $\Delta v = 0.1$ mm, which yields $N_t = 1,416,100$ and $N = 808,600$ (note that $N_t / N = 1.7513$). For this case, the values in the first column of the capacitance matrix obtained by VoxCap with $N_{vx} = N_{vy} = N_{vz} = 5$ and FastCap with $2^{nd}$, $4^{th}$, and $6^{th}$ order multipole expansions are compared in Figs. 9(a)-(c). Furthermore, Fig. 9(c) also shows the normalized charge distribution on the structure obtained by VoxCap. The results obtained by FastCap with $6^{th}$ order of multipole expansion perfectly match with the results obtained by VoxCap. The maximum relative differences between results obtained by VoxCap and FastCap with $2^{nd}$, $4^{th}$, and $6^{th}$ order multipole expansions are 0.30, 0.048, and 0.011, respectively. The VoxCap and the FastCap with $2^{nd}$, $4^{th}$, and $6^{th}$ order multipole expansions obtained the results after 1343, 600, 600, and 259 iterations, respectively. The memory and CPU time requirements of both simulators are tabulated in Table IV. For the same level of accuracy, the VoxCap required 47x less memory and 11.9x less CPU time compared to the FastCap with $6^{th}$ order multipole expansion. For this problem, Tucker enhancement achieves more than 844x reduction (for $tol = 10^{-4}$) in the memory of circulant tensors. Furthermore, it reduces the memory requirement of the simulator from 2.74 GB to 1.92 GB by a factor 1.43 while increasing the CPU time requirement 27%.

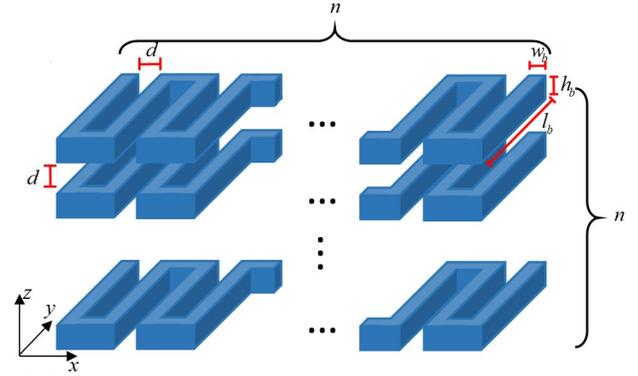

Fig. 8. Parallel meander line structure.

TABLE IV
MEMORY AND CPU TIME REQUIREMENTS OF VOXCAP AND FASTCAP SIMULATORS FOR PARALLEL MEANDER LINE EXAMPLE

| | Memory requirement (MB) | CPU time requirement (s) |
|---|---|---|
| VoxCap (w/out Tucker) | 2,744 | 1,906 |
| VoxCap (w/ Tucker) | 1,918 | 2,426 |
| FastCap w/ $2^{nd}$ order expansion | 22,400 | 5,077 |
| FastCap w/ $4^{th}$ order expansion | 46,800 | 12,246 |
| FastCap w/ $6^{th}$ order expansion | 90,400 | 28,808 |

Finally, a very large-scale parallel meander line structure is formed by setting $w_b = h_b = 0.5$ mm, $d = 0.1$ mm, $l_b = 50$ mm, and $n = 100$. The discretization of this structure by $\Delta v = 0.1$ mm yields $N_t = 179,400,500$ and $N = 100,203,000$ ($N_t / N = 1.79$). For this example, the detailed breakdown of the memory and CPU usage of the VoxCap simulator is provided in Table V. It is apparent in Table V that the Tucker decomposition reduces the memory requirement of the circulant tensors from 132 GB to 6.8627 MB (for $tol = 10^{-4}$); it achieves a CR more than 19,000 while imposing a CO around 0.01. While compressing circulant tensors via Tucker decomposition, the required peak memory is 94.6 GB. Compared to the memory requirement of one circulant tensor 22 GB, the memory penalty is 72.6GB. Furthermore, by reading the Tucker compressed Toeplitz tensors and restoring them, the VoxCap reduces the time for obtaining circulant tensors from 3832.63 seconds to 83.1 seconds; it achieves 46x speed-up. Tucker decomposition reduces the memory requirement (peak memory) from 350 GB to 240 GB by a factor 1.46 while introducing a computational penalty of 25%. When the number of voxels in each small box for the preconditioner is increased from $N_{vx} = N_{vy} = N_{vz} = 10$ to $N_{vx} = N_{vy} = N_{vz} = 20$, the memory requirement of the preconditioner is increased from 320 MB to 19.5 GB, but the solution time is reduced nearly by a factor of 1.8 (from 47,620 secs to 26,773 secs) (Table V). Fig. 9(d) shows the values in the one column of the capacitance matrix obtained by the VoxCap simulator. Note that the FastCap cannot be executed for this large-scale example due to its high memory cost. Fig. 9(d) also presents the normalized charge distribution on the structure by zooming into the opposite corners of the structure.

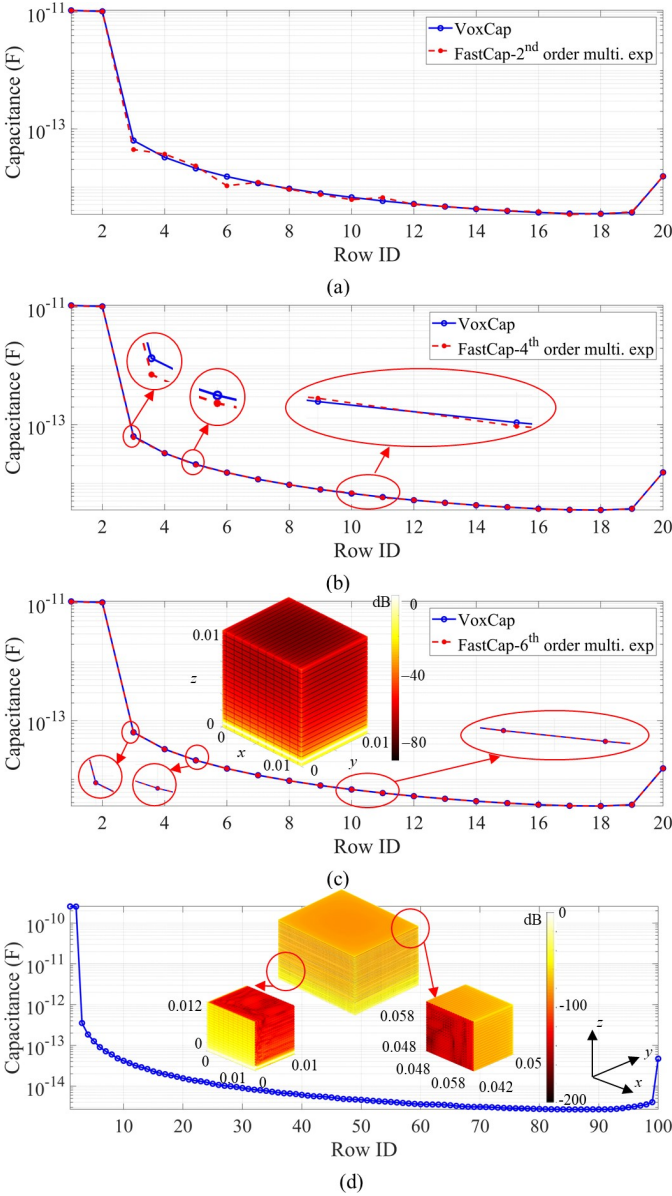

Fig. 9. Parallel meander lines example. For $n = 20$ and $l_b = 10$ mm, the comparison of the values in the first column of the capacitance matrix obtained by the VoxCap and the FastCap with (a) 2nd order and (b) 4th order multipole expansions, as well as (c) 6th order multipole expansions and normalized charge distribution on the structure. (d) For $n = 100$ and $l_b = 50$ mm, the values in the first column of the capacitance matrix obtained by the VoxCap and the normalized charge distribution on the structure and on its left lower and right upper corners.

TABLE V
THE DETAILED BREAKDOWN OF THE CPU TIME AND MEMORY USAGE FOR THE PARALLEL MEANDER LINE. UNITS FOR MEMORY AND CPU TIME ARE MB AND S.

| | |
|---|---|
| CPU time for pre-processing | 1365.8 |
| Memory of original $\tilde{\mathcal{P}}^{\alpha,\beta}$ | 132,099.6 |
| Memory of compressed $\tilde{\mathcal{P}}^{\alpha,\beta}$ | 6.8627 |
| CPU time for reading compressed Toeplitz tensors from hard disk | 0.10129 |
| CPU time for restoring compressed Toeplitz tensors and embedding circulant tensors | 83 |
| CPU time for filling Toeplitz tensors and embedding circulant tensors | 3,832.63 |
| CPU time for compressing the circulant tensors | 1,295.53 |
| CPU time for fast Fourier transforming the circulant tensors | 262.47 |
| CPU time / memory for the preconditioner ($N_{vx} = N_{vy} = N_{vz} = 10$) | 195.33 / 320.27 |
| CPU time / memory for the preconditioner ($N_{vx} = N_{vy} = N_{vz} = 20$) | 245.88 / 19,500.1 |
| Number of iterations / CPU time for the solution when bottom conductor is excited (for $N_{vx} = N_{vy} = N_{vz} = 10$) | 242 / 47,620.00 |
| Number of iterations / CPU time for the solution when bottom conductor is excited (for $N_{vx} = N_{vy} = N_{vz} = 20$) | 157 / 26,773.02 |

## IV. CONCLUSION

In this paper, VoxCap, a Tucker-enhanced and FFT-accelerated SIE simulator for electrostatic analysis and capacitance extraction of voxelized structures, was introduced. The VoxCap solves SIEs after discretizing the charge densities on panels by piecewise constant basis functions, applying Galerkin testing, and obtaining an LSE. The proposed VoxCap simulator uses FFTs to accelerate the MVMs during the iterative solution of LSE. Furthermore, it makes use of a highly effective and memory-efficient block-diagonal-diagonal preconditioner to reduce the number of iterations. It exploits Tucker decompositions to reduce its setup time and memory footprint. The proposed VoxCap simulator can solve problems with hundreds of million unknowns on a desktop computer. For many practical scenarios comprising densely packed interconnects, it is much faster and memory efficient compared to the FastCap. For example, for a parallel bus scenario considered in the numerical example section, the VoxCap is 11x faster than the FastCap. At the same time, the VoxCap requires 23x less memory compared to the FastCap for the same level of accuracy.

## APPENDIX A
### ANALYTICAL EXPRESSION FOR THE RESULT OF INTEGRAL IN (6)

The analytical expression for the result of the integral in (5) is given in Appendix C of [29]. However, no analytical expression for the result of the integral in (6) was found in the literature. To this end, we derived the analytical expression of the result of the integral in (6) by computing the derivative of the analytical expressions in [29] and provide here. For the parallel panel interactions, the result of the integral in (6), $I$, is obtained as

$$I = \sum_{k=1}^{4}\sum_{m=1}^{4}(-1)^{m+k}\left[\frac{0.5a_k z(b_m^2 - z^2)}{(a_k + r_{km})r_{km} + \varepsilon}\right.$$

$$-a_k z \log(a_k + r_{km} + \varepsilon) + \frac{0.5 b_m z(a_k^2 - z^2)}{(b_m + r_{km})r_{km} + \varepsilon}$$

$$-b_m z \log(b_m + r_{km} + \varepsilon) + \frac{2zr_{km}}{3} - z\left(\frac{a_k^2 + b_m^2 - 2z^2}{6r_{km}}\right) \quad (15)$$

$$-a_k b_m \arctan\left(\frac{a_k b_m}{zr_{km}}\right) + \frac{a_k b_m z\left(\frac{a_k b_m}{r_{km}^3} + \frac{a_k b_m}{z^2 r_{km}}\right)}{\frac{a_k^2 b_m^2}{z^2 r_{km}^2} + 1}\right],$$

where $a_1 = x_{s2} - x_{e1}$, $a_2 = x_{e2} - x_{e1}$, $a_3 = x_{e2} - x_{s1}$, $a_4 = x_{s2} - x_{s1}$, $b_1 = y_{s2} - y_{e1}$, $b_2 = y_{e2} - y_{e1}$, $b_3 = y_{e2} - y_{s1}$, $b_4 = y_{s2} - y_{s1}$, $z = z_2 - z_1 + \varepsilon$, $r_{km} = \sqrt{a_k^2 + b_m^2 + z^2}$, and $\varepsilon = 10^{-37}$. The geometrical quantities in these expressions are given in Fig. A(a). For the orthogonal interactions, the result of the integral in (6), $I$, is obtained as

$$I = \sum_{k=1}^{4}\sum_{m=1}^{2}\sum_{l=1}^{2}(-1)^{m+k+l}\left[\frac{b_m(3a_k^2 - b_m^2)}{b_m r_{kml}} + \right.$$

$$\frac{(3a_k^2 c_l^2 - c_l^4) + 3(a_k^2 - c_l^2)(r_{kml}^2 + b_m r_{kml})\log(b_m + r_{kml} + \varepsilon)}{6r_{kml}(b_m + r_{kml} + \varepsilon)}$$

$$+\frac{a_k b_m c_l^2}{r_{kml}(a_k + r_{kml})} + a_k b_m \log(a_k + r_{kml} + \varepsilon) \quad (16)$$

$$-\frac{b_m(r_{kml}^2 + c_l^2)}{3r_{kml}} - \frac{a_k^4 b_m}{6r_{kml}(a_k^2 + c_l^2)} - \frac{a_k^2 b_m^3}{2r_{kml}(b_m^2 + c_l^2)}$$

$$+\frac{a_k c_l}{2}\left(\frac{a_k b_m c_l(r_{kml}^2 + c_l^2)}{r_{kml}(a_k^2 + c_l^2)(b_m^2 + c_l^2)} - 2\arctan(\frac{a_k b_m}{c_l r_{kml}} + \varepsilon))\right],$$

where $a_1 = x_{s2} - x_{e1}$, $a_2 = x_{e2} - x_{e1}$, $a_3 = x_{e2} - x_{s1}$, $a_4 = x_{s2} - x_{s1}$, $b_1 = y_2 - y_{s1}$, $b_2 = y_2 - y_{e1}$, $c_1 = z_{e2} - z_1$, $c_2 = z_{s2} - z_1$, $r_{kml} = \sqrt{a_k^2 + b_m^2 + c_l^2}$, and $\varepsilon = 10^{-37}$. The geometrical quantities in these expressions are provided in Fig. A(b).

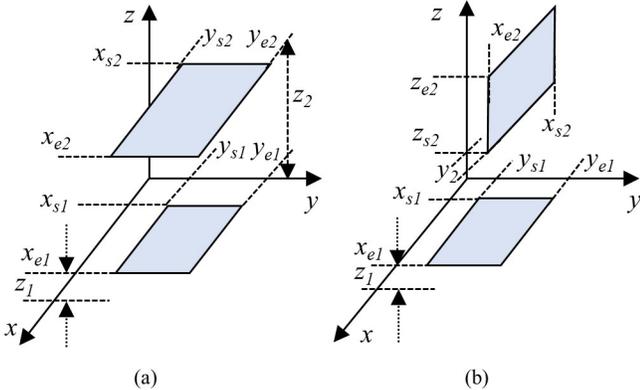

(a) (b)
Fig. A. The sketches show the geometrical quantities used in (15) and (16) for evaluating the interactions between (a) parallel and (b) orthogonal panels.

## APPENDIX B
## BLOCK TOEPLITZ AND CIRCULANT TENSORS

The block Toeplitz tensors $\mathcal{A}^{\alpha,\beta}$ and $\mathcal{B}^{\alpha,\beta}$ corresponding to the block matrices $\bar{\mathbf{P}}^{\alpha,\beta}$ and $\bar{\mathbf{E}}^{\alpha,\beta}$ in (9) are computed to obtain the block circulant tensors $\mathcal{P}^{\alpha,\beta}$ and $\mathcal{E}^{\alpha,\beta}$, respectively. The procedure to obtain $\mathcal{A}^{\alpha,\beta}$ and $\mathcal{B}^{\alpha,\beta}$ is given in Algorithm 1. In this procedure, the source basis function is defined on one of the three orthogonal panels of the first voxel with index $(1,1,1)$, which has unit normal pointing along $x$-, $y$-, or $z$-direction. Next, the testing functions are defined on voxel panels with unit normal pointing along $x$-, $y$- and $z$-directions. By setting the basis function on one orthogonal panel and sweeping over the testing functions, the entries of $\mathcal{A}^{\alpha,\beta}$ and $\mathcal{B}^{\alpha,\beta}$ are obtained via evaluating the integrals in (5) and (6) for given basis function-testing function pair, respectively.

TABLE VI
THE VALUE OF $P_x$, $P_y$, AND $P_z$ FOR DIFFERENT $(\alpha, \beta)$ PAIRS

| $\{\alpha, \beta\}$ | $P_x$ | $P_y$ | $P_z$ |
|---|---|---|---|
| $\{x, x\}$ | $N_x + 1$ | $N_y$ | $N_z$ |
| $\{y, y\}$ | $N_x$ | $N_y + 1$ | $N_z$ |
| $\{z, z\}$ | $N_x$ | $N_y$ | $N_z + 1$ |
| $\{x, y\}, \{y, x\}$ | $N_x + 2$ | $N_y + 2$ | $N_z$ |
| $\{x, z\}, \{z, x\}$ | $N_x + 2$ | $N_y$ | $N_z + 2$ |
| $\{y, z\}, \{z, y\}$ | $N_x$ | $N_y + 2$ | $N_z + 2$ |

TABLE VII
THE VALUE OF $S_x$, $S_y$, AND $S_z$ FOR DIFFERENT $(\alpha, \beta)$ PAIRS

| $\{\alpha, \beta\}$ | $S_x$ | $S_y$ | $S_z$ |
|---|---|---|---|
| $\{x,x\}, \{y,y\}, \{z,z\}$ | 0 | 0 | 0 |
| $\{x, y\}$ | $\Delta v / 2$ | $-\Delta v / 2$ | 0 |
| $\{x, z\}$ | $\Delta v / 2$ | 0 | $-\Delta v / 2$ |
| $\{y, z\}$ | 0 | $\Delta v / 2$ | $-\Delta v / 2$ |
| $\{y, x\}$ | $-\Delta v / 2$ | $\Delta v / 2$ | 0 |
| $\{z, x\}$ | $-\Delta v / 2$ | 0 | $\Delta v / 2$ |
| $\{z, y\}$ | 0 | $-\Delta v / 2$ | $\Delta v / 2$ |

The block circulant tensors $\mathcal{P}^{\alpha,\beta}$ and $\mathcal{E}^{\alpha,\beta}$ are obtained by properly embedding the block Toeplitz tensors $\mathcal{A}^{\alpha,\beta}$ and $\mathcal{B}^{\alpha,\beta}$, respectively. To this end, the procedure explained in the Appendix B of [21] is followed. The signs of the Toeplitz blocks used to construct the blocks of circulant tensors are assigned as in Table VIII. Note that this table corresponds to Table 5 of [21] and the blocks in circulant tensors are labeled by L, M, N, LM, LN, MN, LMN, as in [21].

**Algorithm 1** Procedure for generating $\mathcal{A}^{\alpha,\beta}$ and $\mathcal{B}^{\alpha,\beta}$.

Preprocessing: For given $(\alpha, \beta)$ pair and $\Delta v$,
- Set $P_x$, $P_y$, and $P_z$ by using Table VI and panels' unit normal
- Set the panel center for basis function as $\mathbf{S} = (S_x, S_y, S_z)$ by using Table VII.

Execution:
**for** $m_x = 1 : P_x$ **do**
  **for** $m_y = 1 : P_y$ **do**
    **for** $m_z = 1 : P_z$ **do**
      Set $\mathbf{m} = (m_x, m_y, m_z)$
      Set the panel center for testing function as $\mathbf{O} = (\mathbf{m} - 1)\Delta v$
      Evaluate (5) for given $\mathbf{S}$ and $\mathbf{O}$ to obtain $\mathcal{A}_{\mathbf{m}}^{\alpha,\beta}$ or
      Evaluate (6) for given $\mathbf{S}$ and $\mathbf{O}$ to obtain $\mathcal{B}_{\mathbf{m}}^{\alpha,\beta}$
    **end for**
  **end for**
**end for**

TABLE VIII
THE SIGNS OF BLOCK TOEPLITZ TENSOR FOR CONSTRUCTING EACH BLOCK IN EACH CIRCULANT TENSOR

| Toeplitz tensor \ Block | L | M | N | LM | LN | MN | LMN |
|---|---|---|---|---|---|---|---|
| $\mathcal{A}^{x,x}$, $\mathcal{A}^{y,y}$, $\mathcal{A}^{z,z}$, $\mathcal{A}^{x,y}$, $\mathcal{A}^{x,z}$, or $\mathcal{A}^{y,z}$ | + | + | + | + | + | + | + |
| $\mathcal{B}^{x,x}$, $\mathcal{B}^{x,y}$, $\mathcal{B}^{x,z}$ | - | + | + | - | - | + | - |
| $\mathcal{B}^{y,x}$, $\mathcal{B}^{y,y}$, $\mathcal{B}^{y,z}$ | + | - | + | - | + | - | - |
| $\mathcal{B}^{z,x}$, $\mathcal{B}^{z,y}$, $\mathcal{B}^{z,z}$ | + | + | - | + | - | - | - |

The embedding process generates the block circulant tensors $\mathcal{P}^{\alpha,\beta}$ and $\mathcal{E}^{\alpha,\beta}$ with dimensions given in Table IX. These dimensions are enlarged to $2(N_x+1) \times 2(N_y+1) \times 2(N_z+1)$ by proper zero-padding to reduce the number of FFT and IFFT operations, as mentioned above. The locations of tensor entries for padding zeros are provided in Table X.

TABLE IX
THE ORIGINAL DIMENSIONS OF $\mathcal{P}^{\alpha,\beta}$ AND $\mathcal{E}^{\alpha,\beta}$

| Superscript of tensor $\{\alpha, \beta\}$ | Dimensions | Superscript of tensor $\{\alpha, \beta\}$ | Dimensions |
|---|---|---|---|
| $\{x,x\}$ | $2(N_x+1) \times 2N_y \times 2N_z$ | $\{x,y\}$, $\{y,x\}$ | $2(N_x+1) \times 2(N_y+1) \times 2N_z$ |
| $\{y,y\}$ | $2N_x \times 2(N_y+1) \times 2N_z$ | $\{x,z\}$, $\{z,x\}$ | $2(N_x+1) \times 2N_y \times 2(N_z+1)$ |
| $\{z,z\}$ | $2N_x \times 2N_y \times 2(N_z+1)$ | $\{y,z\}$, $\{z,y\}$ | $2N_x \times 2(N_y+1) \times 2(N_z+1)$ |

TABLE X
LOCATIONS OF ENTRIES IN $\mathcal{P}^{\alpha,\beta}$ AND $\mathcal{E}^{\alpha,\beta}$ FOR ZERO PADDING

| Superscript of tensor, $\{\alpha, \beta\}$ | 1st dimension | 2nd dimension | 3rd dimension |
|---|---|---|---|
| $\{x,x\}$ | - | $(N_y+1):(N_y+2)$ | $(N_z+1):(N_z+2)$ |
| $\{y,y\}$ | $(N_x+1):(N_x+2)$ | - | $(N_z+1):(N_z+2)$ |
| $\{z,z\}$ | $(N_x+1):(N_x+2)$ | $(N_y+1):(N_y+2)$ | - |
| $\{x,y\}$, $\{y,x\}$ | - | - | $(N_z+1):(N_z+2)$ |
| $\{x,z\}$, $\{z,x\}$ | - | $(N_y+1):(N_y+2)$ | - |
| $\{y,z\}$, $\{z,y\}$ | $(N_x+1):(N_x+2)$ | - | - |

## APPENDIX C
### SCALING FACTORS

To use the same Toeplitz tensors $\mathcal{A}^{\alpha,\beta}$ and $\mathcal{B}^{\alpha,\beta}$ stored on the hard disk for given computational domains with different voxel sizes, those are to be computed and stored for unit voxel size $\Delta v = 1$. For the given voxel size of the problem under investigation, the Toeplitz tensors should be multiplied by the scaling factors after those are read from the hard disk. These scaling factors are derived by the change of variables to make the integrals in (5) and (6) independent from $\Delta v$. To this end, consider the parallel panel configuration demonstrated in Fig. A(a). For this configuration, the integral in (5) is written as

$$I = \frac{1}{4\pi\varepsilon_0} \int_{y_{s1}}^{y_{e1}} \int_{x_{s1}}^{x_{e1}} \int_{y_{s2}}^{y_{e2}} \int_{x_{s2}}^{x_{e2}} \frac{dx'dy'dxdy}{\sqrt{(x-x')^2 + (y-y')^2 + (z-z')^2}}, \quad (17)$$

where $z = z_1$ and $z' = z_2$. The integrals on primed coordinates are evaluated on the source panel (top panel in Fig. A(a)) for the basis function while those on the non-primed coordinates are evaluated on the observer panel (bottom panel in Fig. A(a)) for testing function. In the voxelized setting, the source and observer panels sit on grid points with indices $(m'_x, m'_y, m'_z)$ and $(m_x, m_y, m_z)$. To this end, first the bounds are changed as $x_{s2} = m'_x \Delta v$, $x_{e2} = (m'_x+1)\Delta v$, $y_{s2} = m'_y \Delta v$, $y_{e2} = (m'_y+1)\Delta v$, $x_{s1} = m_x \Delta v$, $x_{e1} = (m_x+1)\Delta v$, $y_{s1} = m_y \Delta v$, and $y_{e1} = (m_y+1)\Delta v$. Then the variables and constants are changed as $x' = a'\Delta v$, $y' = b'\Delta v$, $z_1 = c'\Delta v$, $x = a\Delta v$, $y = b\Delta v$, and $z_2 = c\Delta v$ in (17), which yields

$$I = \frac{(\Delta v^3)}{4\pi\varepsilon_0} \int_{m_y}^{m_y+1} \int_{m_x}^{m_x+1} \int_{m'_y}^{m'_y+1} \int_{m'_x}^{m'_x+1} \frac{da'db'dadb}{\sqrt{(a-a')^2 + (b-b')^2 + (c-c')^2}}. \quad (18)$$

Apparently, the integral of (18) is independent of voxel size and thereby scaling factor is $\Delta v^3$ for the parallel panel interactions. Similarly, the scaling factor is obtained as $\Delta v^3$ for the orthogonal panel interactions by applying the same procedure. A similar procedure is applied for the integral in (6). The scaling factor $\Delta v^2$ is obtained for the parallel and orthogonal panel interactions.

## REFERENCES

TMTT-2020-06-0637 14

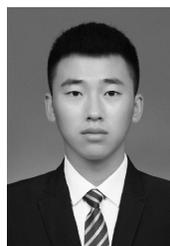

**Mingyu Wang** received the B.S. degree in electrical engineering and automation from Northeast Forestry University, China, in 2016 and M.S. in electronics from Nanyang Technological University, Singapore, in 2018. He is currently working toward the Ph.D. degree at the school of Electrical and Electronic Engineering, Nanyang Technological University, Singapore and works in Computational Electromagnetics Group.

His research is focused on computational electromagnetics and fast parameter extraction of voxelized interconnects and circuits.

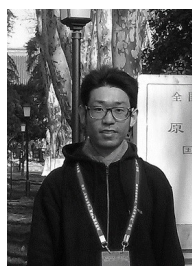

**Cheng Qian** received the B.S. and Ph.D. degree in electronics engineering from Nanjing University of Science and Technology, Jiangsu, China, in 2009 and 2015. From 2016 to 2018, he was a Research Associate with the Department of Applied Physics, The Hong Kong Polytechnic University, Hong Kong. Since 2018, he has been a Post-Doctoral Researcher with the School of Electrical and Electronic Engineering, Nanyang Technological University, Singapore. His current research interests include computational electromagnetics and nonlinear plasmonics.

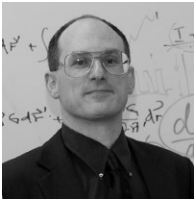

**Jacob K. White** (F'08) received his B.S. in Electrical Engineering and Computer Science (EECS) from the Massachusetts Institute of Technology (MIT) in 1980, and his S.M. and Ph. D. in EECS from the University of California, Berkeley in 1983 and 1985, respectively.

After two years at the IBM T. J. Watson research center, he joined EECS at MIT as an Analog Devices Career Development Assistant Professor in 1987, became a Presidential Young Investigator in 1988, was an associate editor for the IEEE Transactions on Computer-Aided Design from 1992 until 1996, was a member of the Spectre/SpectreRF development team from 1989 until 1999, chaired the International Conference on Computer-Aided Design in 1999, served as an Associate Director of MIT's Research Laboratory of Electronics from 2001 until 2006, was an academic research fellow at Ansoft/Ansys from 2010 until 2016, and served as the MIT EECS co-education officer from 2011 until 2014. He became an IEEE fellow in 2008 for his group's work on fast interconnect analysis (e.g. FastHenry), and shared the 2013 A. R. Newton Technical Impact Award in EDA with Keith Nabors for their fast capacitance extraction program FastCap. Jacob White is currently the C. H. Green Professor in EECS at MIT, where he is researching simulation and optimization techniques for problems in medical technology, nano-photonics, and electrical circuits and interconnect; and experimenting with blended computation- and maker-centric strategies for teaching control, machine learning, and electromagnetics.

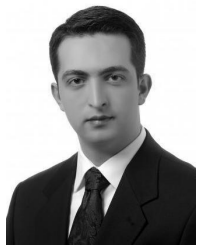

**Abdulkadir C. Yucel** (M'19-SM'20) received the B.S. degree in electronics engineering (*Summa Cum Laude*) from Gebze Institute of Technology, Kocaeli, Turkey, in 2005, and the M.S. and Ph.D. degrees in electrical engineering from the University of Michigan, Ann Arbor, MI, USA, in 2008 and 2013, respectively.

From September 2005 to August 2006, he worked as a Research and Teaching Assistant at Gebze Institute of Technology. From August 2006 to April 2013, he was a Graduate Student Research Assistant at the University of Michigan. Between May 2013 and December 2017, he worked as a Postdoctoral Research Fellow at the University of Michigan, Massachusetts Institute of Technology, and King Abdullah University of Science and Technology. Since 2018, he has been working as an Assistant Professor at the School of Electrical and Electronic Engineering, Nanyang Technological University, Singapore.

Dr. Yucel received the Fulbright Fellowship in 2006, Electrical Engineering and Computer Science Departmental Fellowship of the University of Michigan in 2007, and Student Paper Competition Honorable Mention Award at IEEE AP-S in 2009. He has been serving as an Associate Editor for the International Journal of Numerical Modelling: Electronic Networks, Devices and Fields and as a reviewer for various technical journals. His research interests include various aspects of computational electromagnetics with emphasis on analytical and numerical electromagnetic modelling and the applications of uncertainty quantification and deep/machine learning techniques to the electromagnetic analyses.